\documentclass[preprint,12pt]{elsarticle}

\usepackage{graphics}
\usepackage{graphicx}
\usepackage{hyperref}

\let\stdcaption\caption
\usepackage{float}
\let\caption\stdcaption

\usepackage{amssymb}
\usepackage[latin1]{inputenc} 
\begin{document}

\begin{frontmatter}

%% Title, authors and addresses

%% use the tnoteref command within \title for footnotes;
%% use the tnotetext command for the associated footnote;
%% use the fnref command within \author or \address for footnotes;
%% use the fntext command for the associated footnote;
%% use the corref command within \author for corresponding author footnotes;
%% use the cortext command for the associated footnote;
%% use the ead command for the email address,
%% and the form \ead[url] for the home page:

%\title{Annealing of ZnO surfaces via photoluminescence and molecular dynamics simulation with reactive force field}

\title{Molecular dynamics simulations and photoluminescence measurements of annealed ZnO surfaces}

\author[Label3]{Tjun Kit Min }
\ead{klorin2002@yahoo.co.uk}

\author[Label3]{Tiem Leong Yoon \corref{cor1}}
\ead{tlyoon@usm.my}

\author[Label3]{Chuo Ann Ling}
\ead{lingchuoann@hotmail.com}

\author[Label3]{Shahrom Mahmud}
\ead{shahromx@usm.my}

\author[Label4]{Thong Leng Lim}
\ead{tllim@mmu.edu.my}

\author[Label3,Label5]{Kim Guan Saw}
\ead{kgsaw@usm.my}

\cortext[cor1]{Corresponding author}

%\address[Label1]{Adress for Label1}
%\address[Label2]{Adress for Label2}

\address[Label3]{School of Physics, Universiti Sains Malaysia, 11800 USM, Penang, Malaysia}
\address[Label5]{School of Distance Education, Universiti Sains Malaysia, 11800 USM, Penang, Malaysia}
\address[Label4]{Faculty of Engineering and Technology, Multimedia University,
Jalan Ayer Keroh Lama, 75450 Melaka, Malaysia}

\begin{abstract}
%% Text of abstract

The effect of thermal annealing on wurtzite ZnO, 
terminated by two surfaces, (0 0 0 $\bar 1$) (which is oxygen-terminated) and (0 0 0 1) (which is Zn-terminated), 
is investigated via molecular dynamics simulation using reactive force field (ReaxFF). As a result of annealing at a threshold temperature range of 
%We found that upon heating at or beyond a threshold temperature 
700~K $ < T_{\mbox{\small t}} \leq 800$~K, surface oxygen atoms begin to sublimate from the (0 0 0 $\bar 1$) surface, while no atom leaves the (0 0 0 1) surface. The ratio of oxygen leaving the surface increases with temperature $T$ (for $T \geq T_{\mbox{\small t}}$). 
%A photoluminescence (PL) measurement has been carried out to assess the effect of annealing wurtzite ZnO surfaces. 
The relative luminescence intensity of the secondary peak in the photoluminescence (PL) spectra, interpreted as a measurement of amount of vacancies on the sample surfaces, qualitatively agrees with the threshold behavior as found in the MD simulations. Our simulations have also revealed the formation of oxygen dimers on the surface and evolution of partial charge distribution during the annealing process. Our MD simulation based on the ReaxFF is consistent with experimental observations.

\end{abstract}

\begin{keyword}

Molecular Dynamics; Reactive Force Field; Annealing; ZnO surface

%% MSC codes here, in the form: \MSC code \sep code
%% or \MSC[2008] code \sep code (2000 is the default)
\end{keyword}

\end{frontmatter}

%%
%% Start line numbering here if you want
%%
% \linenumbers

%% main text
\section{Introduction}
\label{intro}

ZnO has been extensively studied, both theoretically and experimentally, due to its many promising applications in piezoelectric devices, transistors, photodiodes, photocatalysis and antibacterial function \cite{Goldberger:JPCB05, Wang:Science06, Becker:JPCC11}. The physical properties of ZnO, especially its surface properties, can be experimentally modified at the atomic level for the purpose of, e.g., engineering the material for desired functionality. Since ZnO contacts with its external environment through its surfaces, monitoring how the surface properties respond to external perturbation (e.g. thermal treatment) can provide valuable insights to improve our ability to manipulate ZnO for application purposes. One of the simplest way to modify the surfaces of ZnO is by heating it to high temperatures (below its melting point). Heating ZnO can be easily carried out in practice, and many works had been reported along this line \cite{Shim:MSEB03,Wei:MC10, Quang:JCG06}.

ZnO crystals are dominated by four surfaces with low Miller indices: the non-polar (1 0 $\bar1$ 0) and (1 1 $\bar2$ 0) surfaces and the polar surfaces which are the zinc-terminated surface (0 0 0 1) and the oxygen terminated surface (0 0 0 $\bar1$). Surface energy of polar surfaces in an ionic model diverges with sample size due to the generation of a macroscopic electrostatic field across the crystal \cite{Kresse:PRB03}.  %The polar surface of ionic crystals will have a surface energy that diverges with sample size due to the generation of a macroscopic electrostatic field across the crystal \cite{Kresse:PRB03}, or the cleavage energy is infinite in the ionic crystals. 
This kind of behavior was well investigated by Tasker \cite{Tasker:JPC79}. This is why wurzite ZnO is labeled as Tusker-type surfaces, which are formed by alternating layers of oppositely charged ions.

In this work we 
%consider specifically the (0 0 0 1) surface of a ZnO crystal in wurtzite structure and question 
investigate what will happen to the atomic configuration of the surface when a ZnO slab with finite thickness is heated without melting. Sublimation of atoms from the polar surfaces due to temperature effect
%, driven by temperature effect, 
%is 
was studied using molecular dynamics (MD) simulation, where the trajectories of all atoms at a given temperature $T$ are followed quantitatively. In the MD simulated annealing, in which the reactive force field (ReaxFF) for ZnO is used, sublimation of O atoms from ZnO polar surface is observed. 
%The occurrence of sublimation is attributed to the novelty of 
ReaxFF for ZnO allows for bond formation and charge transfer among the %participating 
selected atoms. When the sublimation of atoms occurs, point vacancies are created on the surface. Quantitative information of the amount and type of atoms sublimated, as well as point vacancies created on the surface at different annealing temperatures can thus be obtained. 

Experimentally, if we heat a ZnO wurtzite surface to an elevated temperature and investigate the resultant surface using photoluminescence (PL) measurement, the spectrum should reflect the amount of point vacancies created. We expect that an  increase of annealing temperature will create  more point vacancies. In this paper, the predictions of MD simulation using ReaxFF are compared with the PL data.
%to investigate the effects of annealing the wurtzite ZnO surfaces. 

In Section \ref{methods} we describe the MD procedures used to perform the simulations. In Section \ref{experiment} we describe the experimental procedure to perform PL measurement on the ZnO samples. In Section \ref{RnD}, the experimental data and simulation results are discussed, compared and interpreted. The conclusion is in Section \ref{conclusion}.

\section{Molecular dynamics simulation procedure}\label{methods}

\begin{figure}[!ht]
\begin{center}
%\begin{tabular}{p{3in} p{3in}}
\begin{tabular}{c c}
\includegraphics[width=3.2in,height=2in]{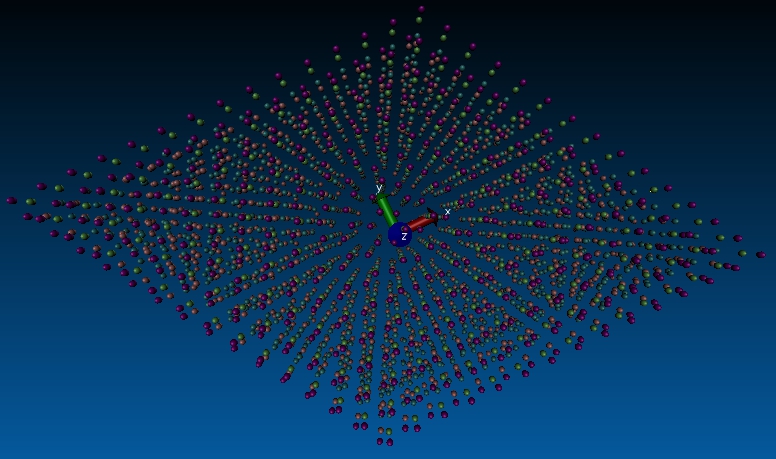} & \includegraphics[width=3.2in,height=2in]{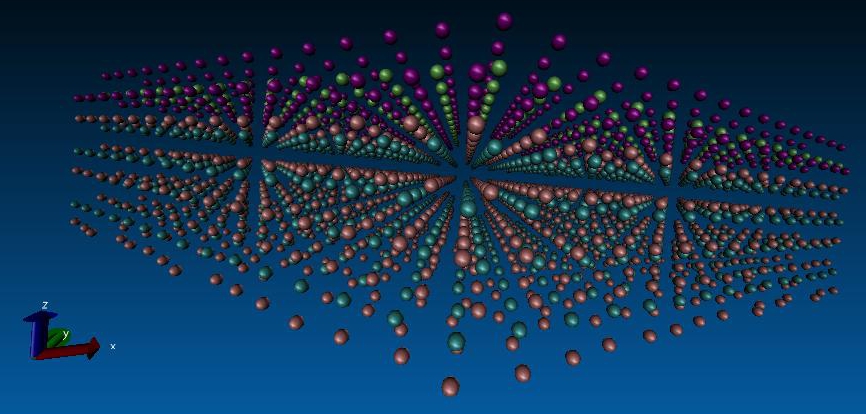}
\end{tabular}
\end{center}
\caption{The 15 $\times$ 15 $\times$ 1 supercell of wurtzite ZnO slab used as initial structure in our MD simulation. Left: Direct surface view from the direction +$z$; Right: Edge-on view. The (0 0 0 $\bar 1$) surface terminates with oxygen atoms while (0 0 0 1) Zn atoms. In these figures, the (0 0 0 $\bar 1$) surface is in the direction pointing along +$z$, while the (0 0 0 1) surface in the -$z$ direction.}
\label{ZnOorig}
\end{figure}

Crystalline ZnO can exist in various polymorphs. The stability of the polymorphs is dependant upon the pressure and temperature \cite{Morkoc:Wiley09}. These polymorphs include wurtzite, zinc blende and rocksalt. ZnO in wurtzite structure is the most stable form at room pressure. The parameters of crystal structure of the wurtzite ZnO unit cell were shown in Table~\ref{ZnOpara}. A slab comprising 15 $\times$ 15 $\times$ 3 unit cells was constructed. The slab was sandwiched between two vacuum layers of thickness 100~$\mbox{\AA}$, while the thickness of the slab itself was 15~\AA. The surface area of the slab in the supercell was 1783~\AA$^2$. The thickness of vacuum was chosen such that the interaction between adjacent surfaces of two neighbouring supercells were negligible so as not give rise to any significant effect on the simulation outcome. Periodic boundary condition was imposed in all $x$-, $y$- and $z$-directions. Fig.~\ref{ZnOorig} shows the supercell of the ZnO slab. In all simulations the total number of atoms in the simulation box were maintained at 2700. 
% By construction, the (0 0 0 1) surface terminated with oxygen atoms while (0 0 0 $\bar 1$) by Zn atoms. 
%To facilitate data abstraction for analysis from the MD results, the atoms in the simulation box are labeled as Zn(1), O(2), Zn(3), O(4) respectively. O(4) refers to oxygen atoms on the outermost (0 0 0 1) surface. Zn(3) labels Zn atoms underneath the O(4) layer. Zn(1) and O(2) lable the rest of the atoms in the slab. 
The MD simulation throughout this work was carried out using the MD code LAMMPS \cite{Plimpton:JCP95}.

The stepsize throughout the simulation was $\Delta t$ = 0.5 fs. The structure was first optimized at 0.1~K using the built-in conjugate gradient minimizer available in the LAMMPS package. We set the convergence criterion to be 10000 steps  (generally, convergence is achieved in less than 100 steps). The temperature of the system was then heated up to $T_{\mbox{\small r}} = 300$~K in 5000 steps. The system was further equilibrated for 20000 steps at 300~K. At the end of the equilibration at $T_{\mbox{\small r}}$, the temperature was raised to a chosen target temperature, $T$, at a rate of 5$\times 10^{10}$ K/s. 
%The target temperature shall be referred as the annealing temperature hereafter. 
The evolution of the system at annealing temperature was then followed at constant $T$ for a total of $N_{\mbox{\small step}}$ = 250000 steps. 
%We shall refer to this phase in the temperature vs. step  profile as the ``temperature plateau'' hereafter. 
The temperature of the system was then quenched from $T$ to 0.1~K at a rate of $5 \times 10^{10}$ K/s. A typical temperature vs. step profile is shown in Fig.~\ref{temperature_vs_I} for annealing temperature $T$ = 1000 K. Nose-Hoover thermostat (NVT) was used throughout the simulation to control the temperature. The damping constant for the thermostat was set to 5 fs. The total steps in each simulation are dependent upon the target temperature $T$. The total duration for the simulations range from 315 000 steps (0.16 ns) to 387 000 steps (0.20 ns).

%We subject a fresh copy of ZnO slab (with initial positions of atoms the same as that in Fig.~\ref{ZnOorig}) to a MD annealing history as that shown in Fig.~\ref{temperature_vs_I}, at a fixed target temperature $T$. 

Independent simulations were also carried out for each target temperature $T$, ranging from 300~K to 1300~K at an interval of 100~K. Atoms from the polar surfaces will leave the surface if the target temperature is sufficiently high. In this case we have to assure that $N_{\mbox{\small step}}$ that defines the length of the temperature plateau is sufficiently long such that atom will stop leaving the surface even if we continue to equilibrate at that annealing temperature. This precaution is necessary so that the conclusion of the MD outcome remains the same should a larger $N_{\mbox{\small step}}$ is used instead. After some trial-and-error efforts, it was found that a value of $N_{\mbox{\small step}} = 250000$ steps was optimal for the range of temperatures investigated in our work. At the end of the simulation, i.e. when the slab has been quenched to 0.1~K from a given target temperature $T$, the type and total number of atoms leaving the surface were quantified. After running a series of simulation for various $T$, we would have collected sufficient data to plot the ratio of atoms leaving the surface as a function of annealing temperature. 

%The radial distribution function of the atoms sitting on the surface of the slab was calculated for the beginning and the end of the annealing process. The surface energies of the slab at various stages during an annealing process was also calculated. 

The reliability of the MD results depends crucially on the force field used. Force field  based on the ReaxFF model, first proposed in \cite{vanDuin:JPCA01}, allows bond formation and charge transfer to occur among the interacting atoms in the simulation box. In this work, the ReaxFF for ZnO is adopted. The ZnO reactive force field has been applied to the calculation of atomic vibrational mean square amplitudes for bulk wurtzite-ZnO for a temperature up to 600~K and found good agreement with experimental observations \cite{Raymand:SS08}. It has also been applied to study surface growth mechanism for the (0 0 0 1) surface and water molecule adsorption on stepped ZnO surfaces \cite{Raymand:SS10}. The parameters of the ReaxFF used in this work was the same as that used in ~\cite{Raymand:SS10}.

Through ReaxFF the process of equilibration of partial charge distribution among the atoms can be revealed in the MD simulation. In our simulations, all the atoms in the simulation box were assigned a partial charge of zero initially. While total charge of the system remains zero, partial charges among the atoms will redistribute itself as thermal equilibration process progresses. We followed the time evolution of partial charge distribution in the slab when the ZnO is going through the annealing process. The distributions of the partial charge density in the slab at the end of a MD run were obtained by post-processing the MD data. 

%distribution as a function of step for every fixed $T$ in the ZnO slab. The partial charge density distribution, i.e., how partial change density varies at different depth of the slab from its surfaces, were also calculated for the range of target temperature simulated. 

\section{Experimental details of photoluminescence measurements}\label{experiment}

Photoluminescence (PL) measurements were carried out to justify the defect caused by the annealing procedure. Two sets of ZnO powder were used as starting material in this study. They were synthesized using French process \cite{Mahmud:JAC11}, having different major morphological particles: rod (ZnO-1) and plate structures (ZnO-2). The ZnO powder were converted into pellets form using the ZnO agglomeration method, which was introduced in the previous work \cite{Mahmud:JAC11}. The ZnO pellets formed using this technique retained the morphological structure of ZnO. X-ray diffraction (XRD) analysis had been carried out on the ZnO pellets and revealed the wurtzite polycrystalline structure of the ZnO samples (not shown here). 

The ZnO pellets were subsequently annealed in nitrogen rich ambient for an hour with the gas flow regulated at 2.4 L/min. The annealing process was done in temperature manipulated at 573~K -- 1073~K.
The PL measurements were taken 
% The experimental work was conducted 
using the Jobin Yvon HR 800UV system 
% photoluminescence spectroscopy (Jobin Yvon HR 800UV) 
with the HeCd Laser 325 nm as excitation source. 
%was used to study the luminescence properties of the sample and to indicate the types of recombination center for the semiconducting samples. 
The level of defect formation was also determined from the luminescence intensity. 
%The experimental results are to be discussed in subsection \ref{comparison}.

\section{Results and Discussion}\label{RnD}

\subsection{Sublimation of O atoms at and beyond a threshold temperature $T_{\mbox{\small t}}$}\label{sublimation}

We have simulated the annealing of the wurtzite ZnO slab from  $T=300$~K to $T=1300$~K at a 100~K interval. At low temperatures, the atoms are observed to execute relatively mild thermal vibration about their equilibrated positions. At higher temperatures, the vibration become more violent and the configuration of the atoms on both surfaces became more distorted as compared to the low temperature cases. However, the layered structure formed by the alternating layers of oxygen and Zn atoms remains visibly recognizable. %As an illustration, Fig.~\ref{3600} displays the snapshots of the ZnO slab at 300, 600 and 1000~K respectively. 

The video of the simulations were visually monitored for any atoms leaving the surfaces of the ZnO slab. We observed that O atoms began to leave the (0 0 0 $\bar 1$) surface, mostly in pairs, for simulation running with $T$ = 800~K and above. Sublimation of the O atoms occurs only during the temperature plateau phase. No atoms other than the O atoms on the (0 0 0 $\bar 1$) surface sublimates from the slab. In particular, no Zn atom sublimates from either surfaces of ZnO. Such observation tallies with the experimental measurement that only oxygen vacancies were created but not Zn upon heating a ZnO sample (see the related discussion in subsection~\ref{comparison}).

\subsection{Ratio of surface O atoms sublimated as a function of annealing temperature}\label{ratioofsurface}
\begin{figure}[!ht]
\begin{center}
\includegraphics[width=3.9in,height=1.70in]{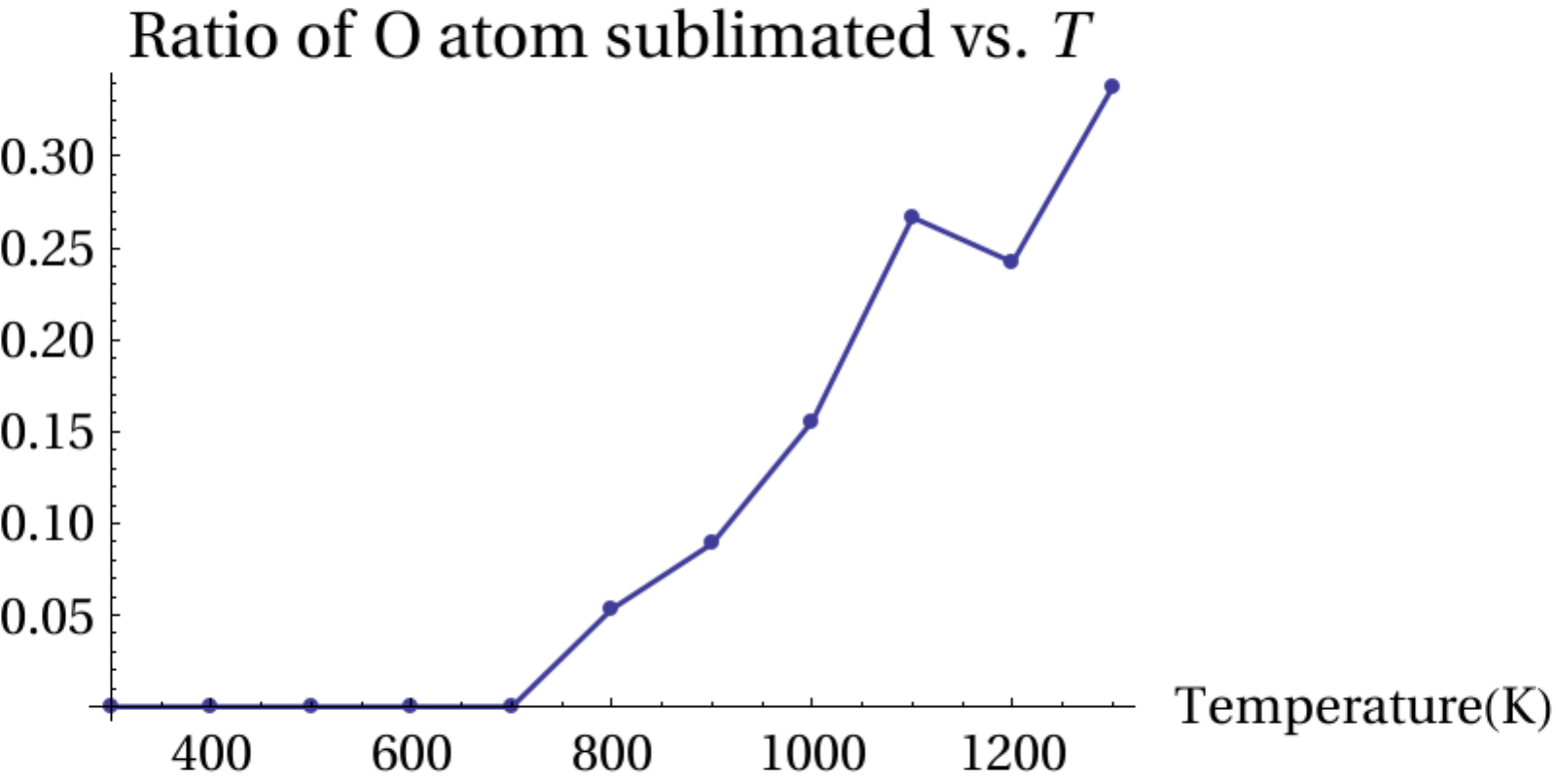}

\end{center}
\caption{Ratio of O atoms sublimated (normalized to original number of  O atoms on the surface) for $T = $ 300 to 1300 K.} 
\label{a4Vovera4VS_vs_temp}
\end{figure}

%%%
%At the end of each simulation for a given $T$, the total number of O atoms sublimated could be easily calculated from the simulation data. 
Fig.~\ref{a4Vovera4VS_vs_temp} shows the ratio of O atoms sublimated to the total O atoms originally lying on the (0 0 0 $\bar 1$) surface for each target temperature. No sublimation occurs for $T = 300, 400, \cdots 700$ K. O atoms sublimate from its polar surface only happened for $T = 800$~K or above, with even more atoms leaving the (0 0 0 $\bar 1$) surface at larger $T$. 
However,  
Fig.~\ref{a4Vovera4VS_vs_temp} 
seems to suggest the possible existence 
%implies the existence 
of a threshold temperature, 700~K $< T_{\mbox{\small t}} \leq 800$~K, at which sublimation of O atoms from the ZnO surface begins to occur. We shall qualitatively compare this prediction against experimentally measured data in subsection \ref{comparison}. 

\subsection{Sublimation of surface O atoms in pairs}
It was also observed that most of the sublimated O atoms will leave the (0 0 0 $\bar 1$) surface in pairs 
%(see Fig.~\ref{3600}~(c) for a graphical illustration)
. A closer investigation of the simulated snapshots revealed that the two oxygen atoms forming a pair during sublimation were neighbors to each other before both left the surface. However we also observed a tiny portion of sublimated O atoms leaving the surface alone and did not form pairs. 

The formation of oxygen atom pairs during sublimation was made possible in the MD simulation due to the novel capability of the reactive force field to allow bond formation between neighboring O atoms. From theoretical point of view, surface atoms are under very different conditions than that for the bulk atoms, give rise to the presence of dangling bonds in the former. As a result, the polar surface which consists of O atoms have higher tendency to form bond with a neighboring O atoms from the same surface layer. When sublimated, these dimers form oxygen pairs as observed in the  MD simulation. 
%The formation of dimers by neighboring O atoms on the surface was similar to dimer formation observed in Si surface~\footnote{For a brief discussion on formation of dimers on the Si surface, see for example, Chapter 1.2 of Ref.~\cite{Mitin:CUP10}.}. 
Based on our MD results, we conclude the following microscopic picture: Dimers of O atoms will be formed on the wurtzite (0 0 0 $\bar 1$) ZnO surface when it is heated up. These dimers shall leave the surface in pairs if subsequently sublimated. 

\subsection{Comparison with results from experiment measurements}\label{comparison}

\begin{figure}[!ht]
\begin{center}
\includegraphics[width=3in,height=4.5in]{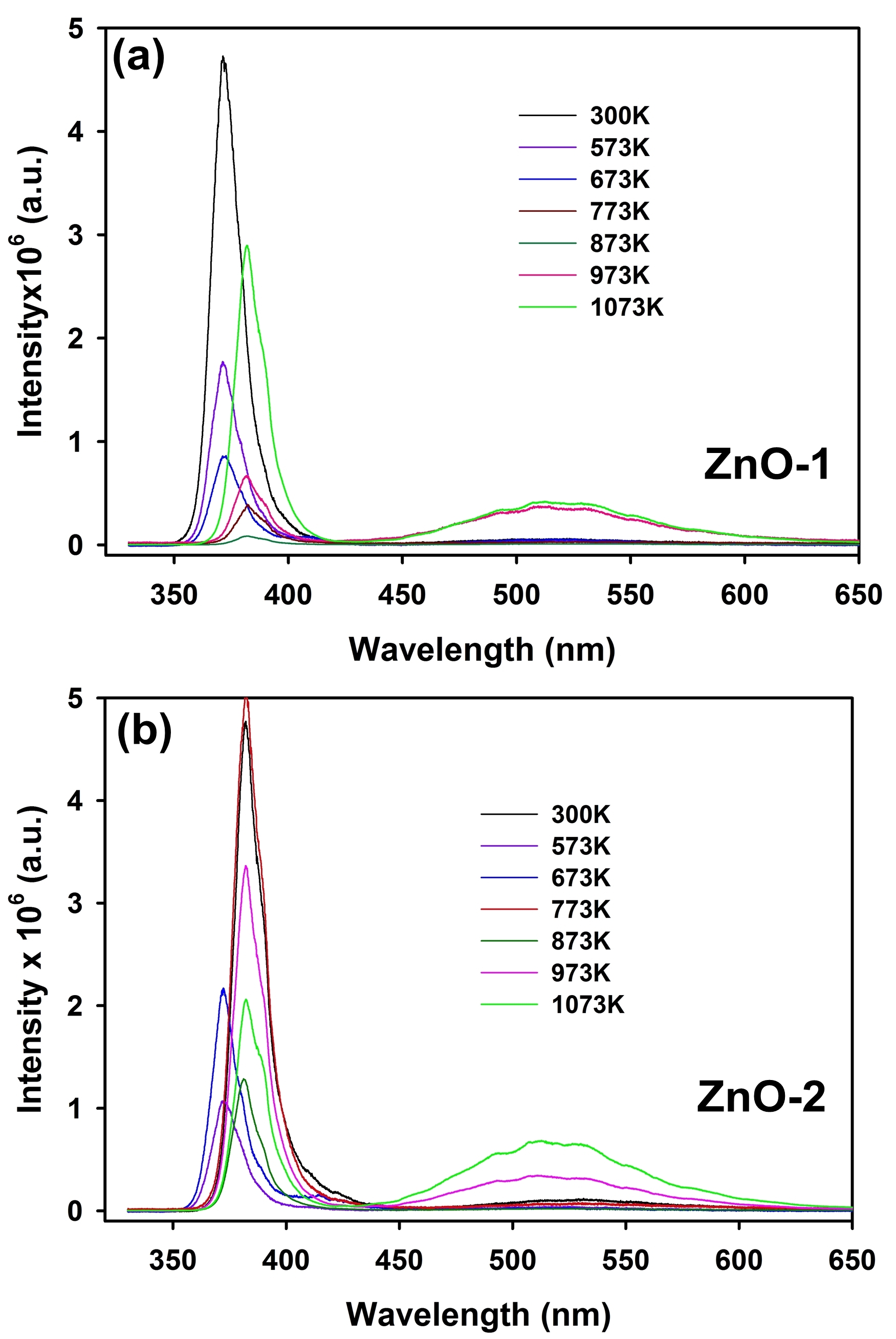}
\end{center}
\caption{PL spectra for two wurtzite ZnO samples thermally treated at various temperatures. The intensity of the secondary peak in the spectrum first appears for $T = $ 973~K. At higher temperatures, the secondary peak records a higher intensity.} 
\label{PLspectrum}
\end{figure}

The experimental data obtained from PL measurement as described in Section~\ref{experiment} is shown in Fig.~\ref{PLspectrum}. The origin of the primary peak (at wavelength 382 nm) in the PL data is due to the bandgap edge as well as excitonic recombination. The origin of the broad PL secondary peak, i.e. green emission centered at 536 nm, remains a debatable issue. It could be interpreted as a reliable indicator of the abundance of point defect present on the surface or otherwise. In either case, it is reasonable to attribute the intensity of the green emission to the presence of point defects. The only issue is the quantitative extent of the contribution of the point defect apart from other possible causes. In this paper we adopt the viewpoint that the intensity of the secondary PL peak is taken as a qualitative indicator of the abundance of point defect present on the surface. Such a viewpoint is consistent with the other reports \cite{Ahn:JAP09,Alvi:NE11,Djurisic:N07,Tam:JPC06,Lin:APL01}. Ahn et al. had explained that the broad green peak centered at 536 nm is due to band transition from zinc interstitial ($\mathrm{Zn}_i$) to oxygen vacancy (Vo) defect levels in ZnO \cite{Ahn:JAP09}. Alvi et. al. reported the band transition from $\mathrm{Zn}_i$ to Vo level is approximately 2.31 eV (green emission) based on the full potential linear muffin-tin orbital method \cite{Alvi:NE11}.

We will interpret the PL data by making the following assumptions: 
\begin{enumerate}
\item The larger the abundance of oxygen point defects on the surface the higher the intensity of the secondary peak;
\item The primary peak contribution to the oxygen vacancies on the surface is not to be considered in present analysis (which is still qualitative) \footnote{Detailed analysis and interpretation of the primary peaks in the PL spectrum for the ZnO samples measured in present study is to be discussed elsewhere. 
}
\end{enumerate}

%To compare the PL measurement with the MD simulation results, we ideally assume that these two quantities follow a proportional relation, i.e., the larger the abundance of point defect on the surface, the higher is the intensity of the green emission in the corresponding PL spectra of the ZnO sample.  

\begin{figure}[!ht]
\begin{center}
\includegraphics[width=3.5in,height=2.5in]{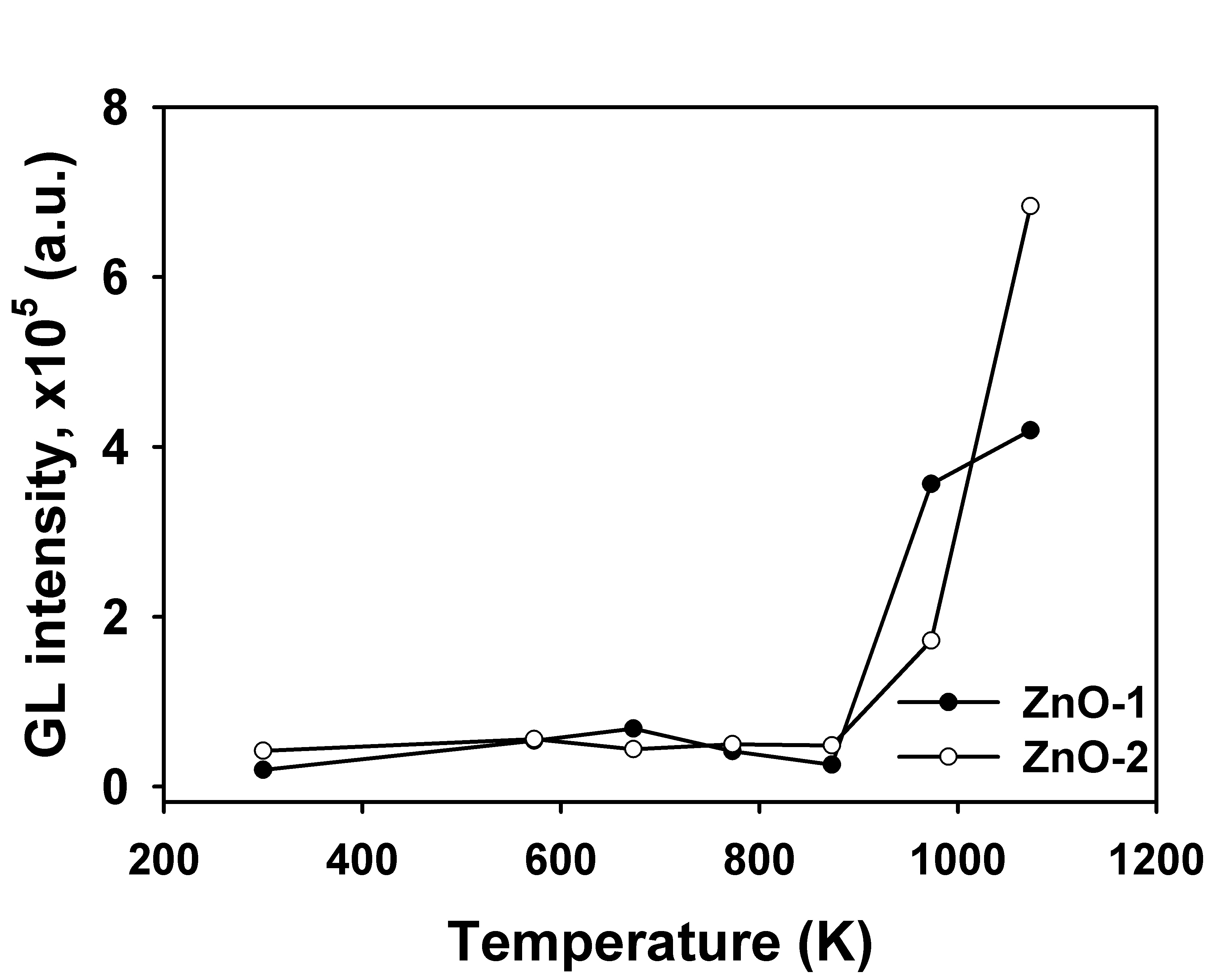}
\end{center}
%\caption{The height of secondary peaks taken at the wavelength 525 nm of Fig.~\ref{PLspectrum} for various annealing temperatures.}
\caption{Green luminescence (512 nm) intensity of the two ZnO samples at different temperatures.} 
\label{PL_vs_temperature}
\end{figure}

Fig.~\ref{PL_vs_temperature} shows the intensities of green luminescence (GL) at wavelength 512 nm of the ZnO-1 and ZnO-2 samples as a function of temperature. The ZnO samples at $T = 300$~K, $T = $ 573 K, 673~K, 773~K and 873~K display a low intensity. There is a pronounced increase of GL luminescence at $T$ = 973~K and $T$ = 1073~K, with the intensity at the latter temperature higher than the former. When $T$ crosses 873~K to 973~K, both ZnO-1 and ZnO-2 samples (with different morphologies) display the same threshold effect of abrupt increase in the GL intensity. A temperature threshold effect is clearly being triggered between $T$ = 873~K and $T$ = 973~K, which is to be compared with $700~\mbox{K} < T_{\mbox{\small t}} \leq $ 800~K predicted by the MD simulation. 

The predicted threshold temperature is based on an idealized model in a MD simulation that cannot exactly describe what actually happened in a real experiment. Since the parametrisation of the ZnO ReaxFF does not include sublimation of O atoms from ZnO surfaces, the value of $T_{\mbox{\small t}}$ from MD falls so closely to the experimentally measured range of $873~\mbox{K} \-- $ 973~K is a very suggestive finding. If we accept an uncertainty of $\sim \pm 100$~K in the threshold temperature, which is reasonably acceptable for a typical MD calculation, then the following picture seems appropriate (at least qualitatively): ({\it i}) Oxygen point defects on the wurtzite ZnO (0 0 0 $\bar 1$) surface are created only at or above the threshold annealing temperature of $T_{\mbox{\small t}}$, and ({\it ii}) the abundance of oxygen point defects on the surface increases as the annealing temperature $T$ increases (where $T \geq T_{\mbox{\small t}}$).  
%Based on the comparison between Fig.~\ref{PL_vs_temperature} against Figs.~\ref{a4Vovera4VS_vs_temp},~\ref{PLspectrum} 
%The two sets of independent results paint a consistent picture on the observation of thermally-driven formation of oxygen vacancies on the ZnO surface. 

\begin{figure}[!ht]
\begin{center}
\includegraphics[width=3.0in,height=2.5in]{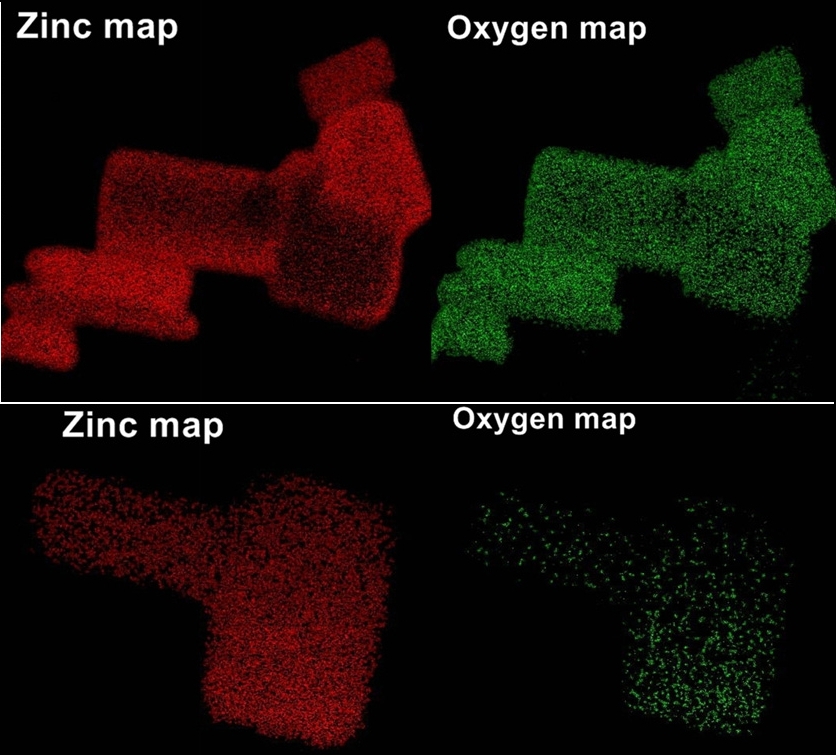}
\caption{ESI-EFTEM images of white ZnO. Above: unannealed sample; Below: $N_2$ annealed (700$^\circ$C) sample. Red dot represents Zn atom while green dot O atom. Images reproduced from Ref.~\cite{Ling:ASS13}.}
\label{fig:Ling}
\end{center}
\end{figure}

%%%
Apart from the PL data as reported above, the MD simulation results are also consistent with a previous experiment \cite{Ling:ASS13} which was  conducted by two present authors (C.A.L and S.M). In \cite{Ling:ASS13}, two different ZnO nanostructures (in wurtzite phase) were investigated using electron spectroscopy imaging (ESI). One sample was not annealed while the other was annealed to $T= 700^\circ $C (= 973 K) under $N_2$-rich condition. 
%Since nitrogen gas is inert, its presence can well be considered to mimic a vacuum condition in the annealing process. 
The surfaces of the samples were scanned for Zn and O atoms. The abundance of these atoms on the surfaces could be visually inspected and were characterized by two distinct colors. The images of the ESI measurement on the white ZnO samples were reproduced in Fig. \ref{fig:Ling}. It was found that the density of Zn atom count is similar for both annealed and unannealed samples. However, the density of the O atoms count for the annealed sample has clearly decreased compared to the unannealed sample. The ESI images are consistent with the notion that O atoms will be sublimated at a temperature $T \geq T_{\mbox{\small{t}}}$, while Zn atoms will not. Although the ZnO samples were measured for only two temperatures (one at ambient and the other at  973 K) in \cite{Ling:ASS13}, the data is ubiquitously consistent with the present MD results.
%%%

The qualitative agreement between the MD results and the two experimental measurements discussed above is encouraging. It provides a motivation to capitalize the MD simulation as a reliable means to derive useful information of the detailed mechanism of the ZnO surface undergoing annealing at the atomistic level, which is otherwise difficult to obtain via experimental approach alone.

\subsection{Partial charge distribution}\label{pq}

\begin{figure}[!ht]
\begin{center}
\includegraphics[width=5.5in,height=4.23in]{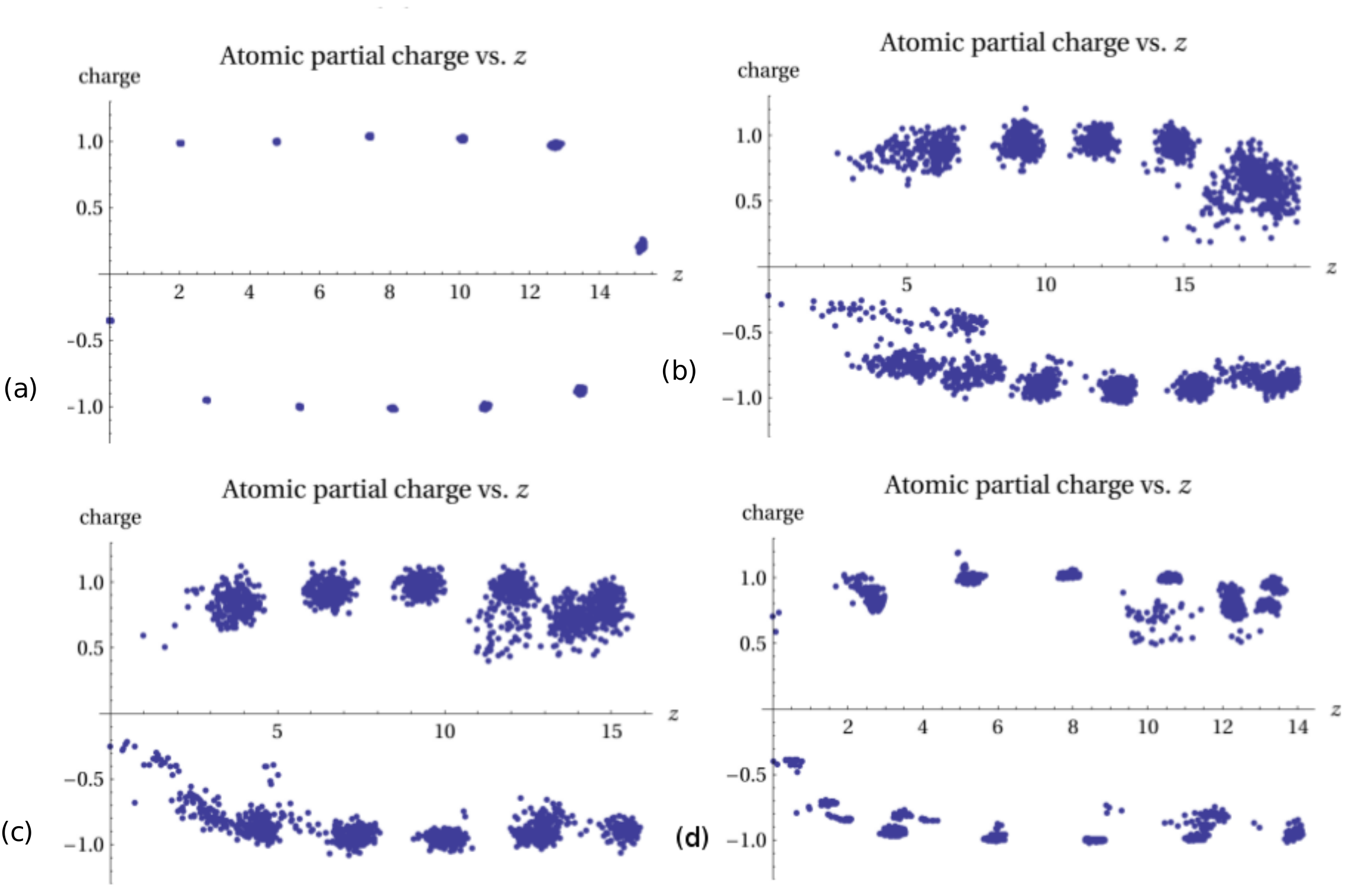}
%\begin{tabular}{p{2.5in} p{2.5in} p{2.5in} }
%\begin{tabular}{c c}
%\begin{tabular}{c c}
%(a) \includegraphics[width=2.6in,height=2in]{partialcharge_vs_z1300step1.pdf} &
%(b) \includegraphics[width=2.6in,height=2in]{partialcharge_vs_z1300step101.pdf} \\
%(c) \includegraphics[width=2.6in,height=2in]{partialcharge_vs_z1300step251.pdf} &
%(d) \includegraphics[width=2.6in,height=2in]{partialcharge_vs_z1300step381.pdf} 
%\end{tabular}
\end{center}
\caption{Snapshots at various stages of the partial charge distribution in the slab when undergoing annealing at $T=$ 1300~K. 
%(a) At the beginning, the slab has only gone through energy minimization at 0.1~K but not any thermal treatment. (b) and (c) are snapshots during which the slab is being annealed at the temperature plateau $T$. (d) Slab after emerged from the end of thermal history as depicted by Fig.~\ref{temperature_vs_I}.
} 
\label{partialcharge_vs_z1300}
\end{figure}

The time evolution of the partial charge distribution among the atoms in the ZnO sample can be followed throughout a MD course (during which the slab is undergoing a temperature history as in Fig.~\ref{temperature_vs_I}). As an illustration, Figs.~\ref{partialcharge_vs_z300}, ~\ref{partialcharge_vs_z1300} display the snapshots of the partial charge distribution in the slab at various steps during the MD simulation at target temperatures $T=$~300~K and $T=$~1300~K, respectively. In these figures, the blobs centered around a given $z$-position (representing the vertical distance from the (0 0 0 $\bar 1$) surface) are made up of discrete dots, each represents the partial charges of an atom at the depth $z$. Partial charge with a positive (negative) value is associated with Zn (oxygen) atom. 

The partial charge distribution for the unannealed slab (i.e. (a) in Fig.~\ref{partialcharge_vs_z300} and Fig.~\ref{partialcharge_vs_z1300}) displays an oscillation pattern. The oscillation is due to the fact that the sign of partial charges from the same atom type at a common layer is opposite to that of adjacent layers occupied by the other atom type. In our MD simulations using ReaxFF model, the partial charge of O atom is negative, while it is positive for Zn atom. The signs are in accordance with what we expected from both atom types based on their known electronegativity. When probing into the depth of the slab along the $z$-direction beginning from the (0 0 0 $\bar 1$) surface termination, atomic layers with alternating partial charges will be passed through in succession, with the first encountered layer being the negatively charged O layer. Oscillation of the partial charge density in the slab observed here has a strong resemblance to what is known as Friedel oscillations \footnote{For a brief discussion of Friedel oscillations, see for example, Chapter 11.2 in the textbook by Kaxiras \cite{Kaxiras:03}.} which lead to surface charge polarization, a common feature of all surfaces.

By following the snapshots sequentially in time, we can quantitative visualize the 
%visually inspect for 
information of how the positions and partial charges of the atoms redistribute themselves throughout a MD course. For example, at the beginning, the blobs of points in (a) in Figs.~\ref{partialcharge_vs_z300}, ~\ref{partialcharge_vs_z1300} are very concentrated. From this, it could be inferred that ({\it i}) all atoms were arranged in layers according to their respective atom type, and ({\it ii}) atoms in the same layer share an approximately same value of partial charges. Atoms in the slab annealed to 300~K vibrate only mildly, as can be inferred from the smaller size of the blobs in (b) and (c) in Fig.~\ref{partialcharge_vs_z300}. At the end of the annealing process at 300~K, the atomic configuration of the slab, as well as the partial charges of the atoms, basically did not alter much. Fig.~\ref{partialcharge_vs_z1300} shows that annealing the slab at 1300~K causes the atoms to 
%for slab annealed at 1300~K, atoms 
vibrate violently about their atomic positions at the temperature plateau (blobs in (b) and (c) are more dispersed than in the 300~K case). At the end of the annealing process, the atomic positions became much distorted. The partial charges of the same atom type in a given layer spread to a broader value (as can be inferred from a large vertical spread in the blobs). Comparing graphs (a) and (d) in Fig.~\ref{partialcharge_vs_z1300}, the distortion resulted from the annealing process at $T = 1300$ K is particularly prominent near the (0 0 0 $\bar 1$) side of the surface, where a large spread in partial charges and locations of the Zn atoms are observed. In general we found that the (0 0 0 $\bar 1$) end of the slab crumpled at the end of the annealing process for $T$ equal or larger than 300~K. 

\begin{figure}[!ht]
\begin{center}
\includegraphics[width=5.5in,height=8.46in]{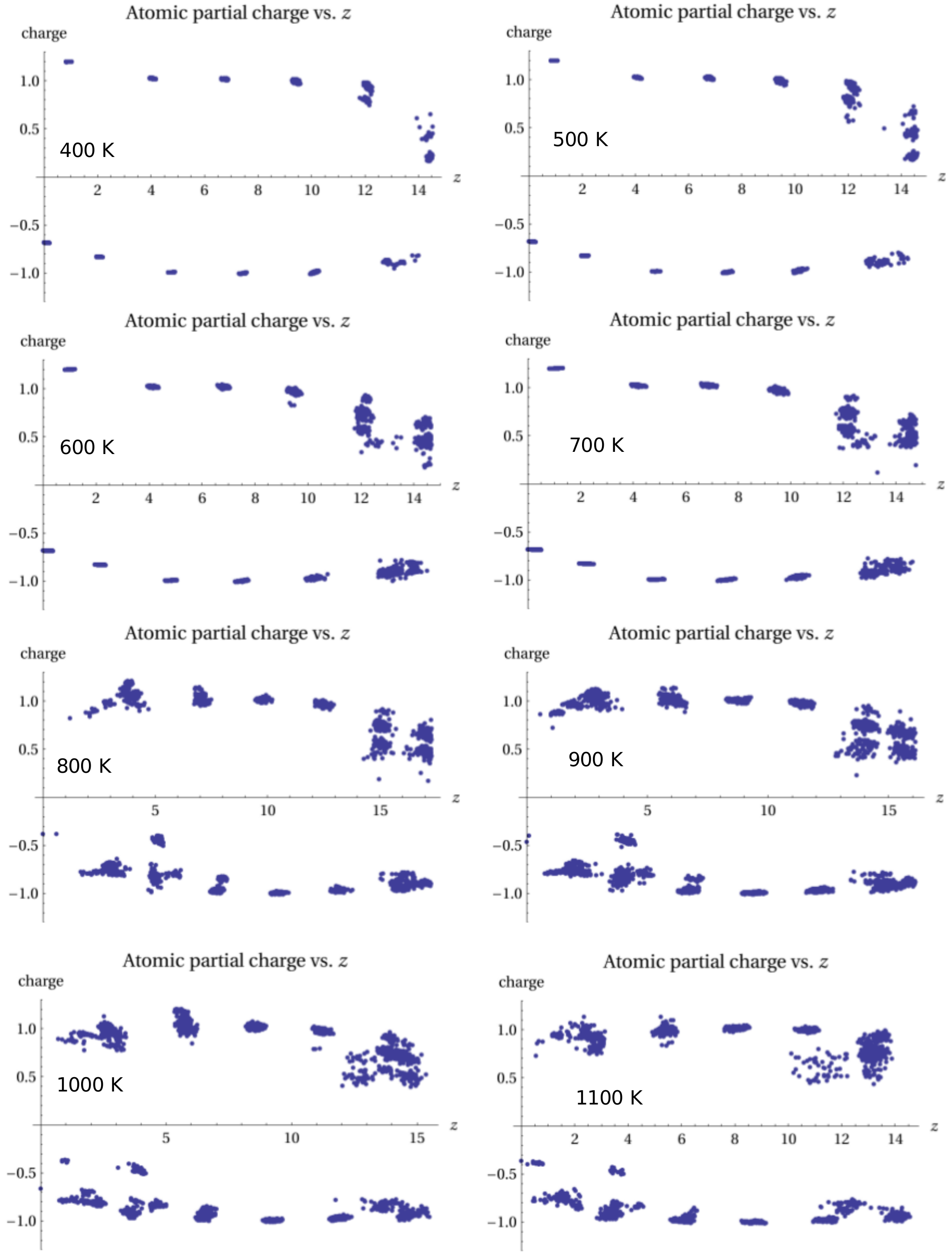}
%\begin{tabular}{p{2.5in} p{2.5in} p{2.5in} }
%\begin{tabular}{c c}
%\begin{tabular}{c c}
%{\footnotesize 400~K} \includegraphics[width=2.5in,height=1.8in]%{partialcharge_vs_z400step311.pdf} &
%{\footnotesize 500~K} \includegraphics[width=2.5in,height=1.8in]%{partialcharge_vs_z500step321.pdf} \\
%{\footnotesize 600~K} \includegraphics[width=2.5in,height=1.8in]{partialcharge_vs_z600step321.pdf} &
%{\footnotesize 700~K} \includegraphics[width=2.5in,height=1.8in]%{partialcharge_vs_z700step331.pdf} \\
%{\footnotesize 800~K} \includegraphics[width=2.5in,height=1.8in]%{partialcharge_vs_z800step341.pdf} &
%{\footnotesize 900~K} \includegraphics[width=2.5in,height=1.8in]%{partialcharge_vs_z900step351.pdf} \\
%{\footnotesize 1000~K} \includegraphics[width=2.5in,height=1.8in]%{partialcharge_vs_z1000step361.pdf} &
%{\footnotesize 1100~K} \includegraphics[width=2.5in,height=1.8in]%{partialcharge_vs_z1100step361.pdf} 
%\end{tabular}
\end{center}
\caption{Partial charge distributions at the end of an MD run for eight  annealing  temperatures ranging from $400$~K to 1100~K.
%Partial charge distributions after the slab emerges from annealing processes at various $T$. 
} 
\label{partialcharge_vs_zlast}
\end{figure}

Fig.~\ref{partialcharge_vs_zlast} displays the partial charge distributions at the end of an MD run for eight  annealing  temperatures ranging from $400$~K to 1100~K. These figures are basically the compilation of stage (d) in Figs.~\ref{partialcharge_vs_z300}, ~\ref{partialcharge_vs_z1300} for eight different $T$ ranging from $400$~K to 1100~K. A qualitative change in the partial charge distribution pattern between $T=$ 700~K and $T=$ 800~K was observed. For $T \leq $ 700~K, the partial charge point were grouped in relatively concentrated blobs, which were slightly flattened in the horizontal direction. Upon crossing the temperature at 800~K, the points of partial charges in these blobs disperse to a much larger extent along the $z$-axis.

\begin{figure}[!ht]
\begin{center}
\includegraphics[width=5.5in,height=4.23in]{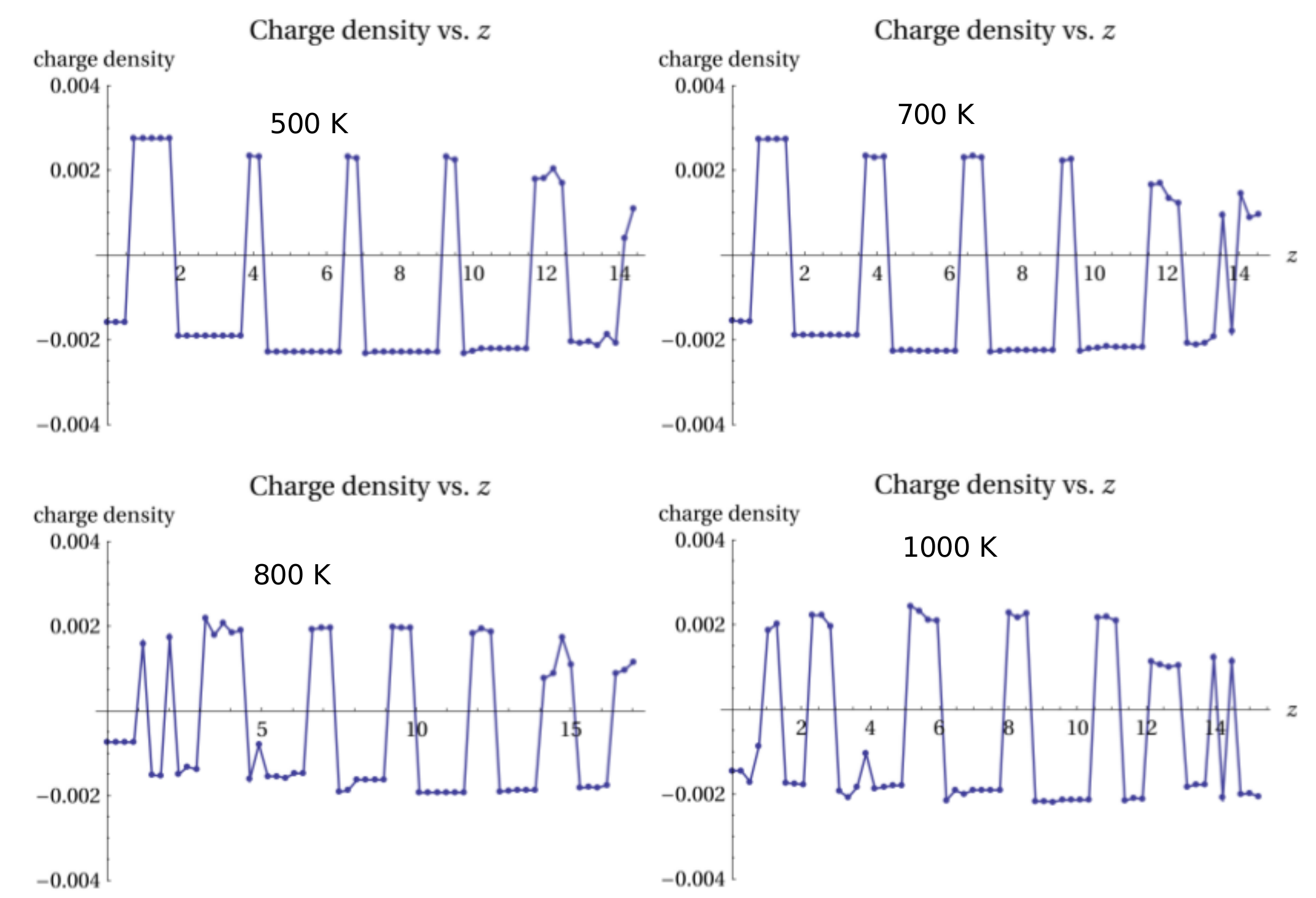}
%\begin{tabular}{p{2.5in} p{2.5in} p{2.5in} }
%\begin{tabular}{c c}
%\begin{tabular}{c c}
%\begin{tabular}{c}\includegraphics[width=2.7in,height=2in]{rho_vs_z500.pdf} \\{\footnotesize $T= 500$~K}\end{tabular} &
%\begin{tabular}{c}\includegraphics[width=2.7in,height=2in]{rho_vs_z700.pdf} \\{\footnotesize $T= 700$~K}  \end{tabular}
%\\
%\\
%\begin{tabular}{c}\includegraphics[width=2.7in,height=2in]{rho_vs_z800.pdf} \\ %{\footnotesize $T= 800$~K }\end{tabular}
%&
%\begin{tabular}{c}\includegraphics[width=2.7in,height=2in]{rho_vs_z1000.pdf} \\ {\footnotesize $T= 1000$~K}\end{tabular}
%\end{tabular}
\end{center}
\caption{Charge density $\rho(z)$ as a function of depth from the (0 0 0 $\bar 1$) surface, $z$, for annealing temperature $T= 500$~K, $T=700$~K, $T=800$~K and $T=1000$~K. The vertical axis is in units of $e/\mbox{\AA}^3$. The $z$-axis is in units of \AA. Note the qualitative change of the density profile (especially the region close to the (0 0 0 $\bar 1$) end) when crossing from $T = 700$~K to $T = 800$~K.
} 
\label{qdensityprofile}
\end{figure}

To illustrate the temperature-induced abrupt change of partial charge distribution from a slightly different perspective, the density of net partial charges at a depth of $z$ from the (0 0 0 $\bar 1$) surface $\rho(z)$, which was derived from the partial charge distribution vs. $z$ data, is shown in Fig.~\ref{qdensityprofile} for annealing temperatures $T = 500, 700, 800, 1000$~K. The density of net partial charges at a depth $z$, $\rho(z)$, is defined as the sum of all partial charges in a volume element of $\Delta z A_s$, where $A_s$ is the surface area of the slab. The variation of $\rho(z)$ at different depth for a few $T$ shown in Fig.~\ref{qdensityprofile} displays a qualitative change in the density of partial charge profile when $T$ crosses 700~K to 800~K. The qualitative change of the partial charge distribution coincides with the onset of O atoms sublimation in this temperature range described in subsection \ref{sublimation}.

\subsection{Partial charge of the sublimated O atoms}

\begin{figure}[!ht]
\begin{center}
\includegraphics[width=4in,height=1.75in]{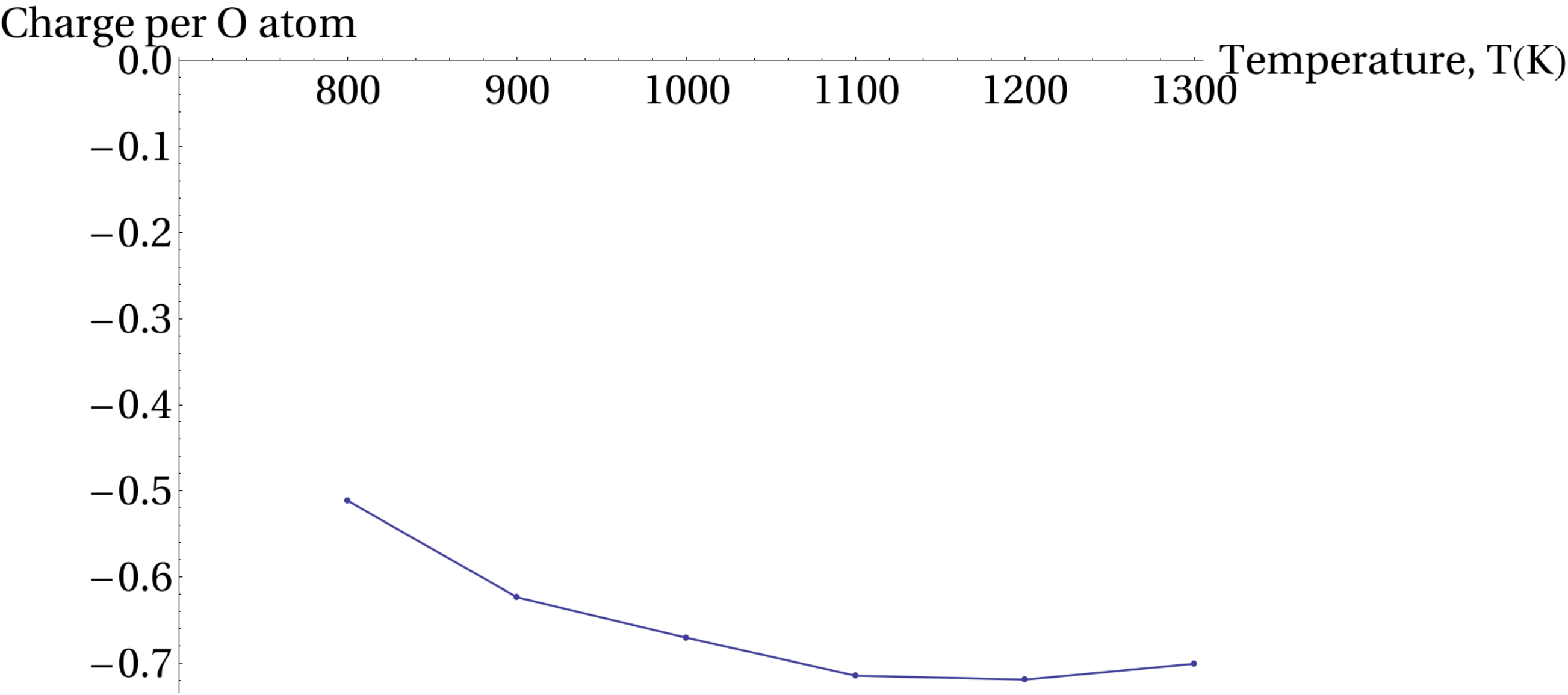}
\caption{Average partial charge per sublimated atom as a function of annealing temperature.}
\label{qV4perAtom_vs_T}
\end{center}
\end{figure}

%In passing, 
We have also enumerated the average partial charge of the sublimated O atoms at each temperature (Fig.~\ref{qV4perAtom_vs_T}). When the sublimated atoms first appeared at $T = T_{\mbox{\small t}}$, the average net charge per atom was $\sim -0.5e$, and this value increased to $\sim -0.7e$ at $T = 1300$~K. The temperature dependence of the average partial charges of the sublimated O atoms appeared to flatten out for $T$ beyond 1100~K. When the O atoms were sublimated from the surface, they carry away with them a net negative charge, leaving behind a substrate that has an overall net positive charge on the surface.

%\subsection{Annealing of alternative ZnO surfaces, (1 0 0 0) and ($\bar {\it 1}$ 0 0 0)}
\subsection{The Zn-terminated surfaces, (0 0 0 1)}

The ZnO slab constructed for the MD simulation contains two surfaces, i.e., (0 0 0 1) and (0 0 0 $\bar 1$). For the Zn-terminated (0 0 0 1) surface, which is on the other side of the slab, 
%These surfaces were obtained by re-orientating the original unit cell. The two surfaces in this construction were terminated by oxygen-zinc hybrid atoms. In fact, most of the low dimensional surfaces of ZnO crystal phases were terminated by Zn or oxygen-zinc hybrid atoms, except the (0 0 0 $\bar 1$) surface in wurtzite phase (which forms the oxygen atom terminated polar surface). The (1 0 0 0) and ($\bar 1$ 0 0 0) are non-polar surfaces.
%In our effort to anneal the (1 0 0 0) and ($\bar 1$ 0 0 0) surfaces of the ZnO slab, 
we have found no robust sign of sublimation of any atom. The  observation 
%The near absence of sublimation 
suggests that O atoms only sublimate from the (0 0 0 $\bar 1$) surface in wurtzite ZnO upon annealing, while not so from the surface terminated by Zn atom. Our MD simulation results suggest the possibility that the Zn atom on the surface termination presents a strong attractive force to any O atom attempting to escape the surface via thermal excitation. Whether the same can be said for other 
%non-polar 
surfaces terminated with Zn- or oxygen-zinc hybrid atoms is an interesting question worthy of further investigation. However such hypothesis needs to be tested more rigorously by investigating the detailed behavior of the force of the Zn atom on any O atom attempting to leave the surface.

\section{Conclusion} \label{conclusion}

\begin{figure}[!ht]
\begin{center}
\includegraphics[width=4.05in,height=1.70in]{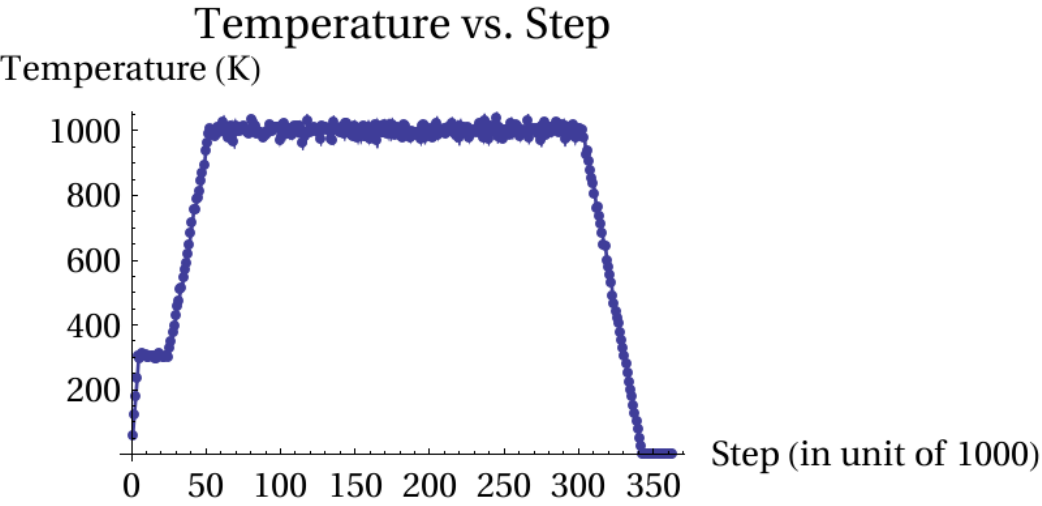}
\end{center}
\caption{A typical temperature vs. step profile in the simulation.}
\label{temperature_vs_I}
\end{figure}

\begin{figure}[!ht]
\begin{center}
%\begin{tabular}{p{2.5in} p{2.5in} p{2.5in} }
%
%\begin{tabular}{c c}
%\includegraphics[width=5.2in,height=4in]{fig7.pdf}
\includegraphics[width=5.5in,height=4.23in]{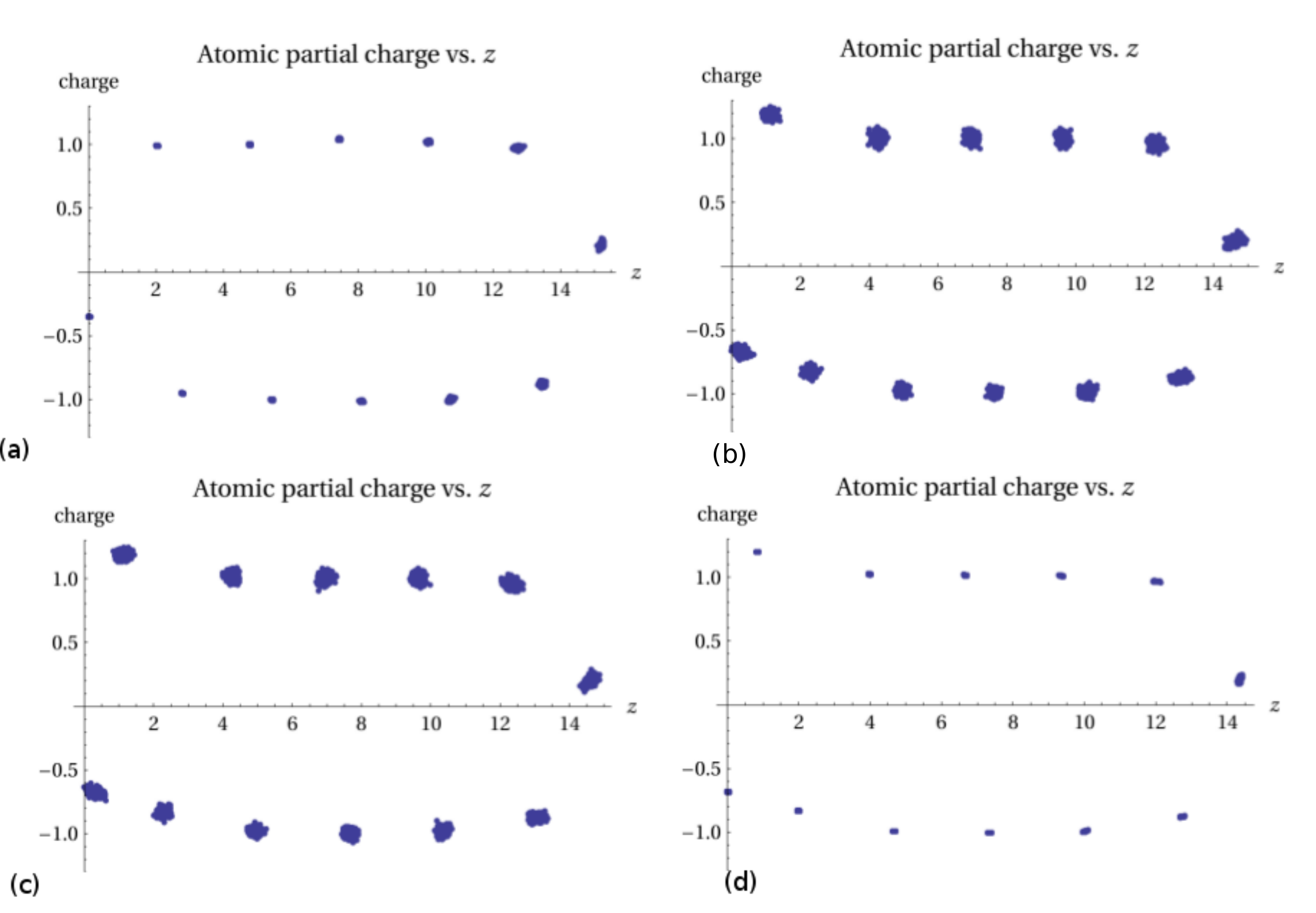}
%\begin{tabular}{c c}
%(a) \includegraphics[width=2.6in,height=2in]{partialcharge_vs_z300step1.pdf} &
%(b) \includegraphics[width=2.6in,height=2in]{partialcharge_vs_z300step101.pdf} \\
%(c) \includegraphics[width=2.6in,height=2in]{partialcharge_vs_z300step251.pdf} &
%(d) \includegraphics[width=2.6in,height=2in]{partialcharge_vs_z300step306.pdf} 
%\end{tabular}
\end{center}

\caption{Snapshots at various stages of the partial charge distribution in the slab when undergoing annealing at $T=$ 300~K. (a) At the beginning, the slab has only gone through energy minimization at 0.1~K but not any thermal treatment. (b) and (c) are snapshots during which the slab is being annealed at the temperature plateau $T$. (d) Slab at
%after emerged from 
the end of thermal history as depicted by Fig.~\ref{temperature_vs_I}. The vertical axis is in units of $e$. The $z$-axis is in units of \AA.}

\label{partialcharge_vs_z300}
\end{figure}

We have performed MD annealing simulations of a wurtzite ZnO slab with a thickness of 15 ${\mbox \AA}$ and a surface area of 1783 \AA$^2$. The simulations are performed using ZnO ReaxFF.
%, as published by \cite{Raymand:SS08, Raymand:SS10}. 
There are two polar surfaces, namely (0 0 0 1) and (0 0 0 $\bar 1$). %The former is O terminated while the latter is Zn terminated. 
The slab is thermally annealed from 0.1~K in stages up to a target temperature $T$, and then further equilibrated at the constant $T$ plateau for sufficiently large number of steps before the temperature is gradually quenched to 0.1 K. The common temperature history experienced by the polar surfaces are shown in Fig.~\ref{temperature_vs_I}.

We have run MD simulations for a range of target temperature from $T = 300$~K till 1300~K at a 100~K interval. Our MD results show that O atoms on the (0 0 0 $\bar 1$) surface are sublimated whenever $T \geq T_{\mbox{\small t}}$, where 700~K $ < T_{\mbox{\small t}} \leq 800$~K. Otherwise no atom sublimates from either surface. The ratio of O atoms sublimated to the total number of O atoms on the surface increases as $T$ ($T \geq T_{\mbox{\small t}}$) increases. The existence of a threshold temperature $T_{\mbox{\small t}}$ at which O atoms from the (0 0 0 $\bar 1$) surface begins to sublimate qualitatively agrees with our PL data (obtained experimentally) and a previous qualitative observation \cite{Ling:ASS13} based on ESI imaging. The MD simulation results also show that the sublimated O atoms do so in pairs. Due to the geometrically asymmetric condition, atoms at the surface are expected to form dimers. The O pairs seen leaving the surface in the MD sublimation  originate from these dimers. We have also investigated the partial charge density of the slab as a function of depth from the surface, $z$, after each slab has gone through an annealing history as depicted in Fig.~\ref{temperature_vs_I}. The oscillatory form of the charge density of the slab resembles Friedel oscillations which are expected for the charge density near a surface. A qualitative change in the partial charge distribution pattern was observed while the  annealing temperature crosses $T=700$~K to $T=800$~K. 
%MD annealing performed on the (1 0 0 0) and ($\bar 1$ 0 0 0) surfaces, which are terminated by oxygen-zinc hybrid atoms, did not produce any sublimation of either O or Zn atoms. 
Thermally-driven sublimation of atoms only occurs in the O-terminated (0 0 0 $\bar 1$) surface, whereas no sublimation of either atom type would occur in the (0 0 0 1) surface which is Zn-terminated, presumably due to a strong attractive force exerted by the Zn atoms at these surfaces.

%The ZnO ReaxFF is a relatively new force field product. Other than the original references, Refs.~\cite{Raymand:SS08,Raymand:SS10}, there has not been many results published using this force field. The MD results as reported here can be taken as a showcase of predictions by the ZnO ReaxFF model.

\section{Acknowledgements}
T. L. Yoon wishes to acknowledge the support of (1) FRGS grant FASA 2/2013 by the Ministry of Higher Education of Malaysia (203/ PFIZIK/6711348), (2) USM RU Grant (1001/PSOSIAL/816210). We also acknowledge the financial support from an ERGS grant (203/PFIZIK/6730100) from the Malaysian Government.

%% References without bibTeX database:
%\bibliographystyle{model1-num-names}
%\bibliography{my}

%\bibliographystyle{unsrt}
%\bibliographystyle{h-physrev3}
\bibliographystyle{model1-num-names}
%\bibliography{../../biblio3}

\newpage
\listoftables
\listoffigures

%%% Tables %%%% 
%\begin{table}[!ht]
%\centering
%\caption{Crystal structure of wurtzite ZnO, as obtained from \cite{urlZnO}}
%\label{ZnOpara} 
%\begin{tabular}{|p{1.2in}|p{0.5in}|p{0.7in}|p{0.7in}|p{0.6in}|p{0.3in}|} \hline 
%    Zn & 1.0  &  0.0   &  0.0     \\ \hline 
%    Zn & 0.5  & 0.2887 &  0.5     \\ \hline 
%    O  & 1.0  &  0.0   & 0.385    \\ \hline 
%    O  & 0.5  & 0.2887 & 0.885    \\ \hline 
%\end{tabular}
%\end{table}
\newpage
\begin{table}[h!]
\centering
\caption{Crystal structure of wurtzite ZnO, as obtained from \cite{urlZnO}}
\label{ZnOpara} 
%\begin{tabular}{|p{1.2in}|p{0.5in}|p{0.7in}|p{0.7in}|p{0.6in}|} \hline 
%Unit Cell Parameters & {Atom} & \textit{x} & \textit{y} & \textit{z}  \\ \hline 
%$ a = 3.3500$ \AA & O & 0.3333 & 0.6667 & 0.3750   \\ \hline 
%{$b = 3.3500$ \AA} & Zn & 0.3333 & 0.6667 & 0.0000   \\ \hline 
%{$c = 5.2200$ \AA}    \\ 
%{$\alpha = 90^\circ$}    \\ 
%{$\beta = 90^\circ$}    \\ 
%{$\gamma = 120^\circ$}  \\ \hline 
%\end{tabular}

%\begin{tabular}{p{1.5in} p{2.5in}} 
%{\begin{tabular}{|p{1.4in}|} \hline 
%Unit Cell Parameters \\ \hline 
%$ a = 3.3500$ \AA    \\ \hline 
%{$b = 3.3500$ \AA}   \\ \hline 
%{$c = 5.2200$ \AA}   \\ \hline 
%{$\alpha = 90^\circ$}\\ \hline 
%{$\beta = 90^\circ$} \\ \hline 
%{$\gamma = 120^\circ$}  \\ \hline 
%\end{tabular}} & 
%{\begin{tabular}{|p{0.5in} p{0.7in} p{0.7in} p{0.6in}|} \hline 
%{Atom} & fractional coordinates \\ \hline 
%{    } & \textit{x} & \textit{y} & \textit{z}  \\ \hline 
% O & 0.3333 & 0.6667 & 0.3750   \\ \hline 
% Zn & 0.3333 & 0.6667 & 0.0000   \\ \hline 
%\end{tabular}}
%\end{tabular}

{\begin{tabular}{|p{0.5in}|p{0.7in}|p{0.7in}|p{0.6in}|} \hline 
\multicolumn{4}{|c|}{Fractional Coordinates of basis atoms} \\ \hline
{Atom} & \textit{x} & \textit{y} & \textit{z}  \\ \hline 
 O & 0.3333 & 0.6667 & 0.3750   \\ \hline 
 Zn & 0.3333 & 0.6667 & 0.0000   \\ \hline 
\end{tabular} 
\begin{tabular}{|p{1.75in}|} \hline 
Unit Cell Parameters \\ \hline 
$ a = 3.3500$ \AA    \\ \hline 
{$b = 3.3500$ \AA}   \\ \hline 
{$c = 5.2200$ \AA}   \\ \hline 
{$\alpha = 90^\circ$}\\ \hline 
{$\beta = 90^\circ$} \\ \hline 
{$\gamma = 120^\circ$}  \\ \hline 
\end{tabular}} 

\end{table}

%%%%% Figures %%%%%%%%%%%%%
\newpage

%\begin{figure}[!ht]
%\begin{center}
%\begin{tabular} {c}
%\begin{tabular} {c c}
%\begin{tabular} {c}
%\includegraphics[width=3.0in,height=1.9in]{300K.jpg} \\ {\footnotesize 300~K}
%\end{tabular}
%& 
%\begin{tabular} {c}
%\includegraphics[width=3.0in,height=1.9in]{600K.jpg} \\ {\footnotesize 600~K}
%\end{tabular}
%\end{tabular}
%\\
%\begin{tabular} {c}
%\includegraphics[width=3.0in,height=1.9in]{1000K.jpg} \\ {\footnotesize 1000~K}
%\end{tabular}
%\end{tabular}
%\end{center}
%\caption{Surface view of the ZnO slab while being annealed at (a) 300~K (b)~600K and (c) 1000~K from the +$z$ direction. In (c) a pair of bonded oxygen atoms being sublimated from the surface is clearly visible on the right of the figure.}
%\label{3600}
%\end{figure}

%%%
%%%

%\begin{figure}[!ht]
%\begin{center}
%\includegraphics[width=5in,height=2.15in]{qVS_vs_i.pdf}
%\caption{An illustration showing the sum of all partial charges in the whole simulation box throughout all steps for MD simulation carried out with target temperature $T = 300$~K. Clearly, the sum fluctuates about the average value of zero. The $y$-axis is in units of $e$. The same curve is obtained for other target temperatures.}
%\label{qVS_vs_i}
%\end{center}
%\end{figure}

\end{document}